\documentclass[12pt,preprint]{aastex}

\shorttitle{\textit{Chandra} Spectroscopy of the Cygnus Loop}
\shortauthors{McEntaffer et al.}

\begin{document}

\title{\textit{Chandra} Imaging and Spectroscopy of the Eastern XA Region of the Cygnus Loop Supernova Remnant}

\author{R. L. McEntaffer, \& T. Brantseg}
\affil{Department of Physics and Astronomy, University of Iowa, Iowa City, IA 52242}
\email{randall-mcentaffer@uiowa.edu}

\begin{abstract}
The XA region of the Cygnus Loop is a bright knot of X-ray emission on the eastern edge of the supernova remnant.  The emission results from the interaction of the supernova blast wave with density enhancements at the edge of a precursor formed cavity.  However, this interaction is complex given the irregular morphology of the cavity wall.  To study the nature and origin of the X-ray emission we use high spatial resolution images from \textit{Chandra}.  We extract spectra from these images to analyze the physical conditions of the plasma.  Our goal is to probe the density of various regions to form a picture of the cavity wall and characterize the interaction between this supernova and the local interstellar medium.  We find that a series of regions along the edge of the X-ray emission appears to trace out the location of the cavity wall.  The best fit plasma models result in two temperature component equilibrium models for each region.  The low temperature components have densities that are an order of magnitude higher than the high temperature components.  The high density plasma may exist in the cavity wall where it equilibrates rapidly and cools efficiently.  The low density plasma is interior to the enhancement and heated further by a reverse shock from the wall.  Calculations of shock velocities and timescales since shock heating are consistent with this interpretation.  Furthermore, we find a bright knot of emission indicative of a discrete interaction of the blast wave with a high density cloud in the cavity wall with a size scale $\sim$0.1 pc.  Aside from this, other extractions made interior to the X-ray edge are confused by line of sight projection of various components.  Some of these regions show evidence of detecting the cavity wall but their location makes the interpretation difficult.  In general, the softer plasmas are well fit at temperatures $\langle kT \rangle \sim$0.11 keV, with harder plasmas at temperatures of $\langle kT \rangle \sim$0.27 keV.  All regions displayed consistent metal depletions most notably in N, O, and Ne at an average of 0.54, 0.55, and 0.36 times solar, respectively.
\end{abstract}


\section{Introduction}\label{intro}
Supernova remnants are a dominant factor in the dynamics of a galaxy given that their ubiquity and size influence all phases of the interstellar medium (ISM).  An investigation of these objects is crucial to understand this interaction and the mechanisms involved with transferring energy and matter to the ISM.  The Cygnus Loop provides an excellent laboratory for the study of a galactic supernova remnant as it is large, $\sim3^{\circ}\times3^{\circ}$ \citep{L97}, close by, 540 pc \citep{blair05}, and relatively unabsorbed ($E(B-V)=0.08$; \cite{Fesen}).  \citet{L97,L98} performed a detailed study of the optical and X-ray emission, and found that the morphology is not indicative of a typical theoretical blast wave in a uniform medium \citep{sedov,spitzer}.  Instead, the precursor wind carved out a nearly spherical cavity made of clumps at high density separated by lower density gas, thus providing the ingredients for a cavity explosion \citep{McCraySnow}.  Over the history of this remnant the blast wave has propagated through a low density plasma and has relatively recently interacted with the cavity wall leading to limb-brightened emission.  The density enhancements decelerate the blast wave and lead to the X-ray emission.

X-ray studies of specific blast wave-cloud interactions give mixed results.  Several studies suggest low temperature ($kT\sim$0.1--0.2 keV) equilibrium conditions within these dense clouds with higher temperatures toward the interior \citep{L02,L05,leahy}.  Alternatively, \citet{Katsuda08} and \citet{M07} find nearly all of the best fit temperatures above $kT=0.2$ keV with nonequilibrium models.  However, the general trend is a multiple temperature component model that is consistent with a lower temperature exterior and higher temperature interior, which may be additionally heated by reverse shocks arising from the cloud interactions.  

The XA region is a complicated, convoluted region of shocks located on the eastern side of the Cygnus Loop, on the south end of NGC 6992 \citep{HesterCox}.  Previous studies have looked at this region over a range of wavelengths \citep{Sankrit,Danforth01,MT01}.  \citet{Sankrit} find complicated ionization structures within UV shocks.  \citet{Danforth01} also perform UV emission line studies and formulate a picture of the XA region where a finger of dense gas is protruding into this region.  In addition, \citet{MT01} observed this region with the \textit{ASCA} and \textit{ROSAT} observatories.  Similar to the above X-ray studies, they find inhomogeneous structure and explain their results with a reflection-shock model resulting from the interaction of the blast wave with a single cavity wall cloud.  More recently, \citet{Zhou} analyzed higher resolution \textit{XMM-Newton} data.  They find a low temperature exterior ($kT\sim$0.07--0.15 keV) with depleted abundances and high temperature interior ($kT\sim$0.24--0.46 keV).  They argue that the region is dense and clumpy, but identify no small scale features.

The goal of this analysis is to exploit the high spatial resolution of \textit{Chandra} to elucidate structure in the XA region and obtain more detail on the interaction of the blast wave with the precursor formed cavity wall.  Combined with the ability to extract spectra, it will be possible to probe the XA region for additional complexities and attempt to model the cavity wall morphology and its physical state.

\section{Observations and Data Reduction}\label{obs}
The bright, shock rich region on the eastern edge of the Cygnus Loop, known as the XA region, was observed with the Advanced CCD Imaging Spectrometer on the \textit{Chandra X-ray Observatory} for 75 ks.  The observation was made on 2001 July 28 and archived under observation ID 1961.  The region was centered on the S3 chip which is back-illuminated and therefore more sensitive to the large amount of soft X-rays being emitted from this region.  This paper concentrates on the interaction of the supernova blast wave with the cavity boundary, which is dominated by soft X-rays.  For that reason we have decided to concentrate our analysis on data contained within only the S3 chip giving a $8.3' \times 8.3'$ field of view.  The observation produced imaging with a spatial resolution of $\sim1''$ and spectral resolution E/$\Delta$E$\sim$5--20 over an energy band of E$\sim$0.5--1.5 keV (flux above 1.5 keV is negligible).

The level 1 data were reprocessed to level 2 with standard processing procedures in the Chandra Interactive Analysis of Observations (CIAO, v4.1.2, \citet{ciao}) software package with current calibration data from the Chandra Calibration Database (CALDB, v.4.1.0).  Good time intervals, charge transfer efficiency, and time dependent gain variations were accounted for.  The data were background corrected using extractions from eastern areas of the chip that are devoid of emission.

\begin{figure} [htbp]
   \centering
%   \plotone{f1.eps}
   \includegraphics[width=6.0in,height=5.46in]{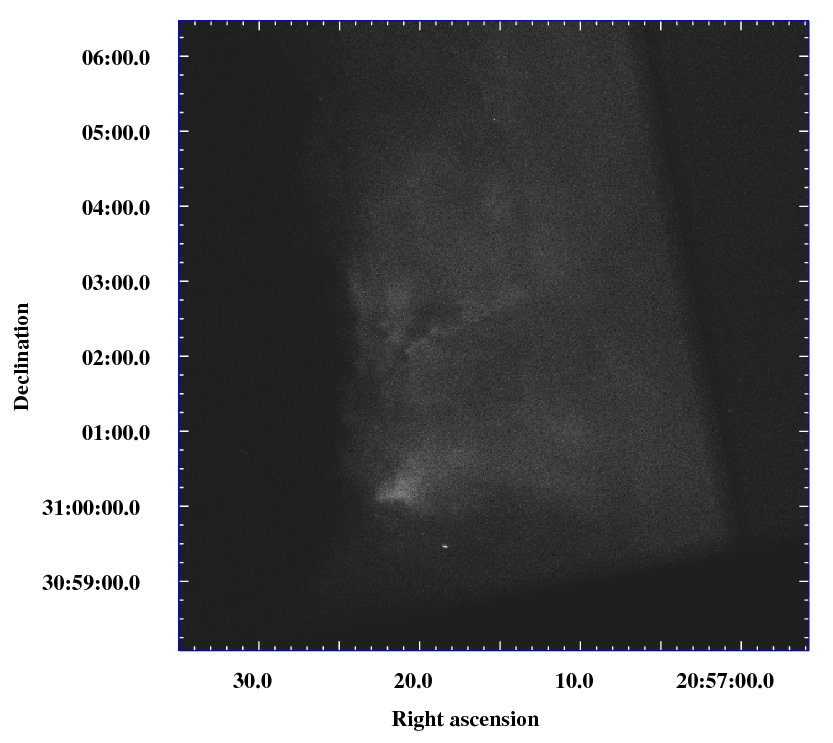} 
   \caption{Image of the raw data from the \textit{Chandra} ACIS S3 back-illuminated CCD.  The pixels have been binned by 2 and are shown linearly scaled with values ranging from 0-70 counts per bin.}
   \label{rawdata}
\end{figure}

\begin{figure} [htbp]
   \centering
%   \plotone{f2.eps}
   \includegraphics[width=6.0in,height=5.29in]{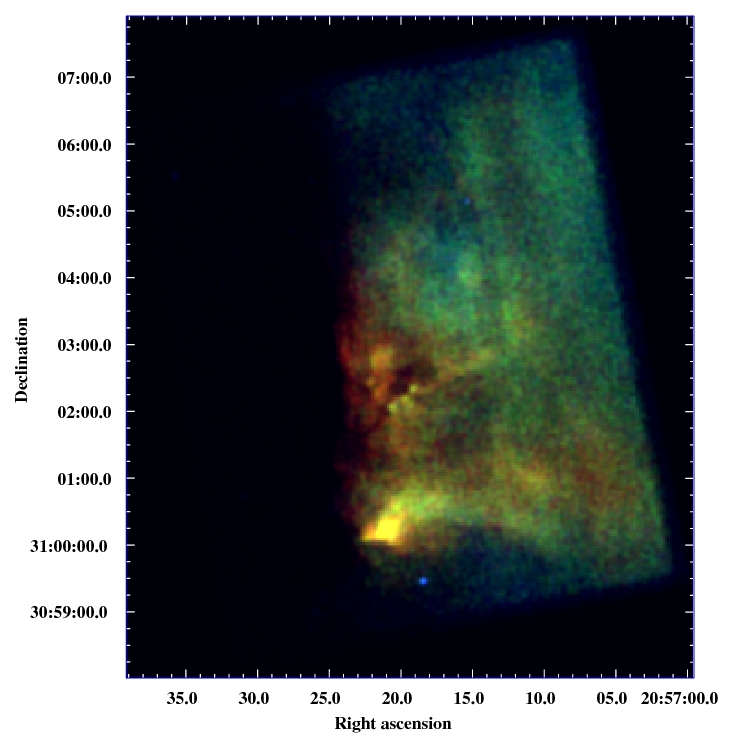} 
   \caption{A false color image of the raw data.  The three colors, red, green, and blue, depict the total counts in three energy bands, 0.3-0.5 keV, 0.5-0.8 keV, and 0.8-1.5 keV, respectively.  The direction of shock propagation is from right to left so that the eastern red edge is closest to the current shock position.}
   \label{color}
\end{figure}

\begin{figure} [htbp]
   \centering
%   \plotone{f3.eps}
   \includegraphics[width=6.0in,height=5.36in]{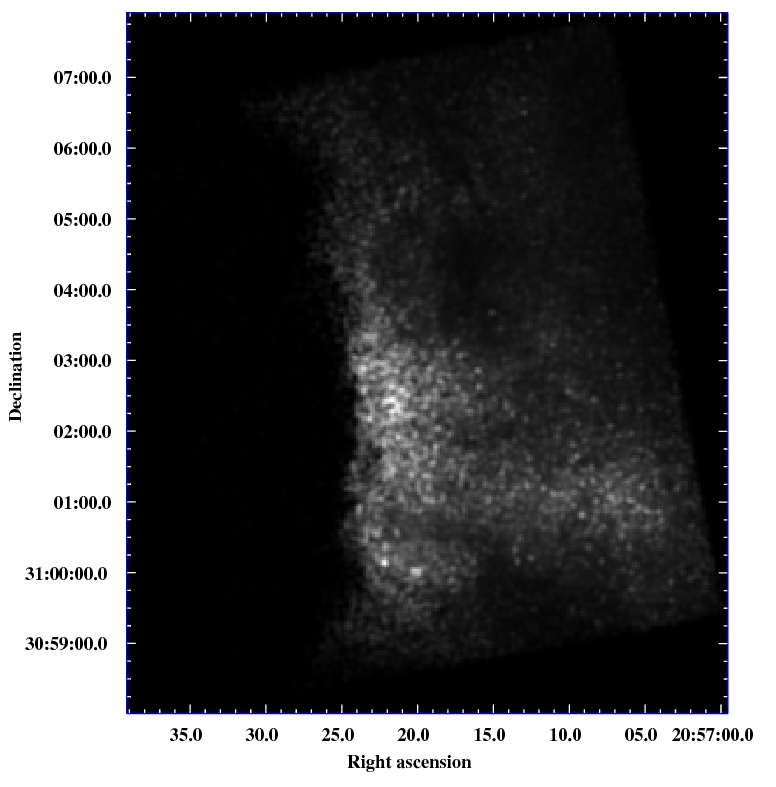} 
   \caption{The ratio of detected soft energy flux (0.3--0.5 keV) to hard energy flux(0.8--1.5 keV).  The bright regions trace the soft X-ray emission arising from the blast wave encountering the local ISM.}
   \label{ratio}
\end{figure}

The reprocessed raw data is displayed in figure~\ref{rawdata}.  It is evident that this region is extremely complex with many shocks existing down to the smallest scales.  The emission appears to be concentrated in clumps or hot spots as opposed to filamentary shock structures.  To elucidate the spectral structure within this region, a false color image, created by discriminating the data via energy, is shown in figure~\ref{color}.  The raw data are segmented into 0.3--0.5 keV (red), 0.5--0.8 keV (green), and 0.8--1.5 keV energy bands.  The soft, red emission is dominant on the very eastern edge where the blast wave is initially encountering the cavity wall, although some red emission is seen toward the interior, especially on the south end.  The harder, green emission is interior to this red emission as is the hardest, blue emission.  To gain more insight into the distribution of these energy regions a ratio map of the soft emission (0.3--0.5 keV) to hard (0.8--1.5 keV) emission is displayed in figure~\ref{ratio}.  This shows the predominance of soft emission at the eastern edge where the blast wave is interacting with the cavity wall.  Also, there is a large, bright, soft region centrally located in declination, along the front, that extends into the remnant considerably with relatively little high energy contribution.  Finally, there is a line of soft emission that extends from the shock front westward into the interior of the remnant along a declination of $\sim31^\circ 01'$.

\section{Spectral Extractions and Fitting}\label{specext}
We extracted data from specific regions in order to perform spectral analyses.  These regions were chosen for their apparent physical qualities based on inspection of the false color image and the band ratio map.  The combination of these images can be used to elucidate regions with distinct physical characteristics.  Spectrally extracted regions are shown in figure~\ref{regions} overlaid on both the false color image and the band ratio map.  Spectra from these regions were created using the \textit{specextract} tool in CIAO.  This tool also creates specific ancillary response files and redistribution matrices for each region.

The first region of interest is the soft X-ray dominated rim of the remnant.  The extraction regions are depicted by the series of rectangles following the rim and labeled B1--B5.  Inspection of the color image shows that these regions are dominated by soft emission. In contrast to these B regions, the regions A1 and A2 are dominated by harder emission with A2 located in a veritable void in the ratio map.  The next region of interest is the centrally located soft X-ray enhancement, probed by extraction regions C1 and C2.  C1 corresponds to a bright knot of emission in the false color image while C2 corresponds to a bright spot in the ratio map.  This area of soft X-ray enhancement appears extremely complex with structures appearing more like clumpy hot spots rather than elongated shock filaments.  Another region of interest lies along the westward soft X-ray extension along declination $\sim31^\circ 01'$.  Regions E1--E3 are used to investigate this emission.  Finally, there is an extremely bright knot in the southeast located at $\sim31^\circ$ of declination and right ascension of 20$^h$ 57$^{min}$ 21$^{sec}$.  The high count rates allow for several extractions over this region.  A grid of extraction boxes is used to investigate the spectral properties of this bright emission.

To analyze the extraction regions we use the Xspec spectral fitting package v12.5.1n \citep{Xspec}.  Only data above 0.5 keV are fit due to the uncertainty of the low energy calibration.  The data are also binned to $>30$ counts per bin so that Gaussian statistics can be applied.  If the X-ray emission arises from supernova blast wave heating of the ISM, then we expect to see cooling emission predominantly in emission lines.  Depending on the density of the plasma and the time since it has been shocked, it will either been in collisional ionization equilibrium (high density or large amount of time since heating) or on the way there, which is described by a nonequilibrium ionization model.  In Xspec, the appropriate models are \textit{vequil} for equilibrium conditions, and \textit{vnei} for nonequilibrium conditions \citep{bork01,bork94,Hamilton,liedahl95}.  These models are convolved with a photoelectic absorption model, \textit{phabs}, which accounts for interstellar absorption along the line of sight.  Dielectronic recombination rates are taken from \citet{mazz} with abundances from \citet{angr} and cross sections from \citet{bcmc}.  The line list used is an improved version of the APED line list \citep{aped1}, which has been developed by Kazik Borkowski and includes more inner shell processes especially for Fe-L lines.  Additional fitting was performed using a plane parallel shock model, \textit{vpshock} \citep{bork01,sedov}, but this never resulted in better fit statistics in comparison to the \textit{vequil} or \textit{vnei} models.

Investigation of emission line lists for plasmas with $kT\sim$0.1--0.2 keV show that the emission is dominated by high ionization states of oxygen, nitrogen, and carbon.  At temperatures of $kT>0.2$ keV iron and neon have significant contributions, especially above photon energies of 0.7 keV.  The fits presented here typically varied the abundances for some subsection of these five elements.  All non-varying elemental abundances are set to solar values.  Allowing $N_H$ to vary causes it to trend toward unphysically low values with large errors in a handful of regions.  Therefore, for all fits the column density of intervening material, $N_H$, was set to $8\times10^{20}$ cm$^{-2}$ consistent with other Cygnus Loop studies \citep{McECash08, Zhou, MT01}.  Other variables include the temperature of the plasma, $kT$, a normalization parameter which includes emission measure, $norm$, and the ionization timescale, $\tau$.  The ionization timescale is the product of the time since shock heating and the electron density of the shocked material, $\tau=n_{e}t$, with $\gtrsim10^{11}$ cm$^{-3}$ s typically signifying equilibrium conditions.  This parameter is only used in the \textit{vnei} models.

A summary of the best fit parameters is shown in Tables~\ref{tbl1}, ~\ref{tbl2}, and ~\ref{tbl3}.  Quoted errors are 90\% confidence intervals for the parameter of interest.  The A regions are best fit with two temperature components, however A1 requires two equilibrium plasmas while A2 uses two plasmas not yet in collisional ionization equilibrium.  Single temperature \textit{vequil} and \textit{vnei} fits were attempted for A1 and A2 but are unable to produce a $\chi^{2}_{red}<2$.  Substituting one of the equilibrium plasmas with a nonequilibrium plasma does not improve the $\chi^2$ statistic for A1 and the $\tau$ parameter is driven to its maximum, thus suggesting equilibrium conditions.  Region A2 could only be adequately fit ($\chi^{2}_{red}<2$) with nonequilibrium plasmas in its two component model.  These fits tie the abundances between C, N, and O together for each component.	In addition to a relatively high $kT$ component, A1 shows the presence of a low temperature plasma at $kT=0.09$ keV while A2 is characterized by only high temperature gas.  The abundances in all components are reduced from solar.

In the B regions, all extracted spectra are best fit with a two temperature equilibrium model.  It is only necessary to vary the abundance of O to obtain an adequate fit.  The abundance of O is tied between the two temperature components.  Allowing C and N to vary does not improve the fit statistics and if allowed to vary their values are typically driven to zero or to unphysically high values, e.g. tens to hundreds.  Again, several combinations of models were attempted to fit each B region.  Single temperature fits (\textit{vpshock}, \textit{vequil}, and \textit{vnei}) were incapable of fitting the data from any of the regions.  In region B1 only a two component \textit{vequil} model is reasonable.  Changing one of the components to nonequilibrium conditions does not improve statistics and the $\tau$ parameter pegs high.  Including a nonequilibrium component in the B2 fit does produce a reasonable fit ($\chi^{2}_{red}=1.3$), but an F-test just favors the two component equilibrium model (F-test value $=3.1$ with probability, $P=0.06$, where $P_{crit}=0.05$).  B3 is similar to B1 in that the inclusion of nonequilibrium conditions does not improve the fit and the $\tau$ parameter reaches its upper limit.  In addition to the two component equilibrium model, the B4 spectrum can be fit by both a \textit{vequil} $+$ \textit{vnei} model and a \textit{vnei} $+$ \textit{vnei} model ($\chi^{2}_{red}=1.6$ for both), but an F-test clearly favors the double equilibrium model (probabilities of 0.34 and 0.45, respectively).  Finally, a \textit{vequil} $+$ \textit{vnei} model fits B5, but with a marginally improved $\chi^{2}_{red}$ leading to an F-test probability of 0.94, thus negating the addition of the nonequilibrium model.  Again, the O abundance is depleted from solar in all regions with a range of 0.51--0.82 (where an abundance of 1.0 would signify solar).  In each region there is a low temperature, soft component with $kT$ ranging from 0.09--0.12 keV, and a harder component with temperatures ranging from 0.26--0.6 keV.

Table~\ref{tbl2} contains the fit parameters for the bright regions encompassed by C1 and C2.  In both cases, the given model is the only reasonable fit obtained for these regions.  The C2 fit is very similar to the B regions - O is the only varied abundance in a two component equilibrium model consisting of a low T and high T plasma ($kT=0.11$ and $0.37$ keV).  The C1 fit is similar to the A2 fit, but at lower temperatures--an average of 0.19 keV versus 0.27 keV.  The abundances of C, N, and O were allowed to vary independently, but each was tied between the two temperature components.  The abundances of N and O are again depleted but there is a suspicious upper limit to C.  Lines from \ion{C}{6} lie below our energy range and would contribute to the shape and trend of the extracted spectrum at low energies.  This high C value is a result of a single high value in the lowest energy bin.  Maintaining a constant level of interstellar absorption leads to an increase in the C abundance to compensate for this high bin.

The E regions sample the soft X-ray enhancement that extends to the west.  The best fit parameters for these regions are given in Table~\ref{tbl3}.  In contrast to the other regions, it is necessary to vary the Ne and Fe abundances in E1--E3 to obtain an adequate fit.  This leads to depletions in O and Ne (0.12--0.25 and 0.32--0.41, respectively) with Fe close to or slightly higher than solar (0.8--1.6).  Region E1 is adequately fit with \textit{vequil} alone as well as a two component \textit{vequil} $+$ \textit{vnei} model, but F-tests performed between these models and \textit{vnei} favor the single temperature nonequilibrium model.  E3 is much different in that only a single temperature equilibrium fit is valid.  Attempts at using \textit{vnei}, alone or in combination, drives the ionization parameter to equilibrium values.  Finally, the only model capable of fitting region E2 is a two temperature component \textit{vequil} $+$ \textit{vnei} model.  The high temperature nonequilibrium component is similar to regions E1 and E3 in temperature and elemental abundances, but the additional equilibrium component is at a lower temperature with solar abundances.

\begin{figure} [htbp]
   \centering
   \epsscale{0.7}
%   \plotone{f5.eps}
   \includegraphics[width=6.5in,height=7.5in]{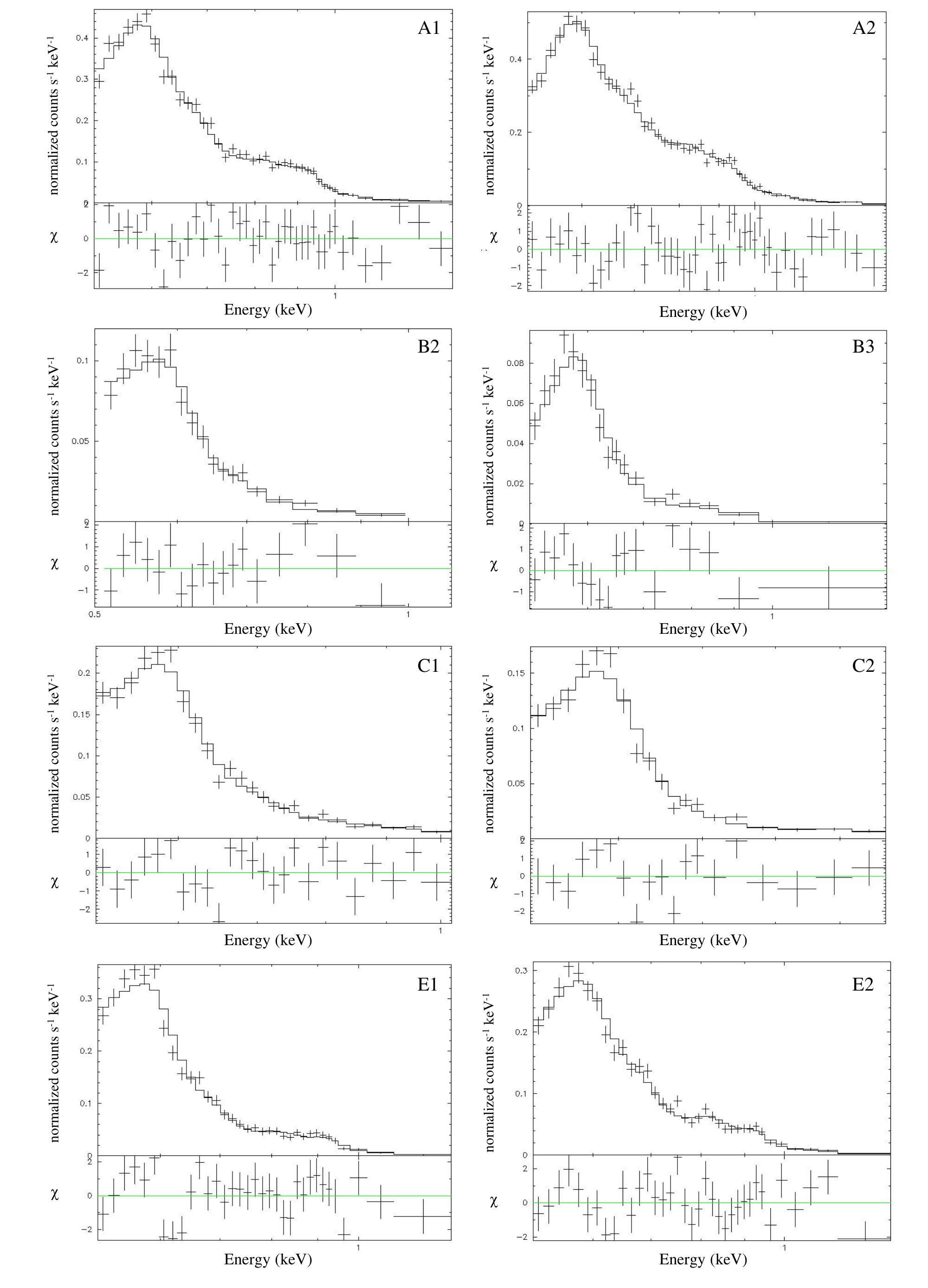}
   \caption{Spectra from the extracted regions with best fit model shown as solid lines.  The A regions are best fit with a nonequilibrium model, the B and C regions with an equilibrium model, and the E regions with a two temperature component model.}
   \label{spectra}
\end{figure}

The extracted spectra from each of the regions discussed above are given in figure~\ref{spectra}.  In addition to the extracted data, the best fit model is shown for each region as a solid line.  Regions B2 and B3 have been chosen as representative spectra from the B regions.  Visual inspection of figure~\ref{spectra} shows that the B and C regions are dominated by soft emission below 700 eV.  More specifically, other than the strong \ion{O}{7} triplet from 561--574 eV, no significant emission lines stand out.  Alternatively, the emission from the A and E regions appears to be harder or at least contains a higher temperature component.  First, in addition to the bright \ion{O}{7} triplet, the presence of \ion{O}{8} at 654 eV, 775 eV, and 817 eV can be inferred.  This may solely be due to the improved count statistics in these bright regions.  More convincingly, there are humps of significant emission above 700 eV present.  This emission is most likely from \ion{Fe}{17} lines which are expected around 725 eV, 727 eV, 739 eV, 812 eV, and 826 eV, as well as the He-like \ion{Ne}{9} triplet around 905--922 eV.  The peaks of emission lines are systematically underestimated which may point to an error or conservatism in the redistribution matrices.

\begin{figure} [htbp]
   \centering
   \epsscale{0.5}
%   \plotone{f6.eps}
   \includegraphics[width=5.0in,height=7.19in]{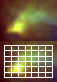}
   \caption{Close-up of the bright knot of emission in the southeast of the S3 CCD field of view.  The bottom image shows an overlay of the extraction region grid.}
   \label{grid}
\end{figure}

\begin{figure} [htbp]
   \centering
   \epsscale{1.0}
%   \plotone{f7.eps}
   \includegraphics[width=6.0in,height=5.14in]{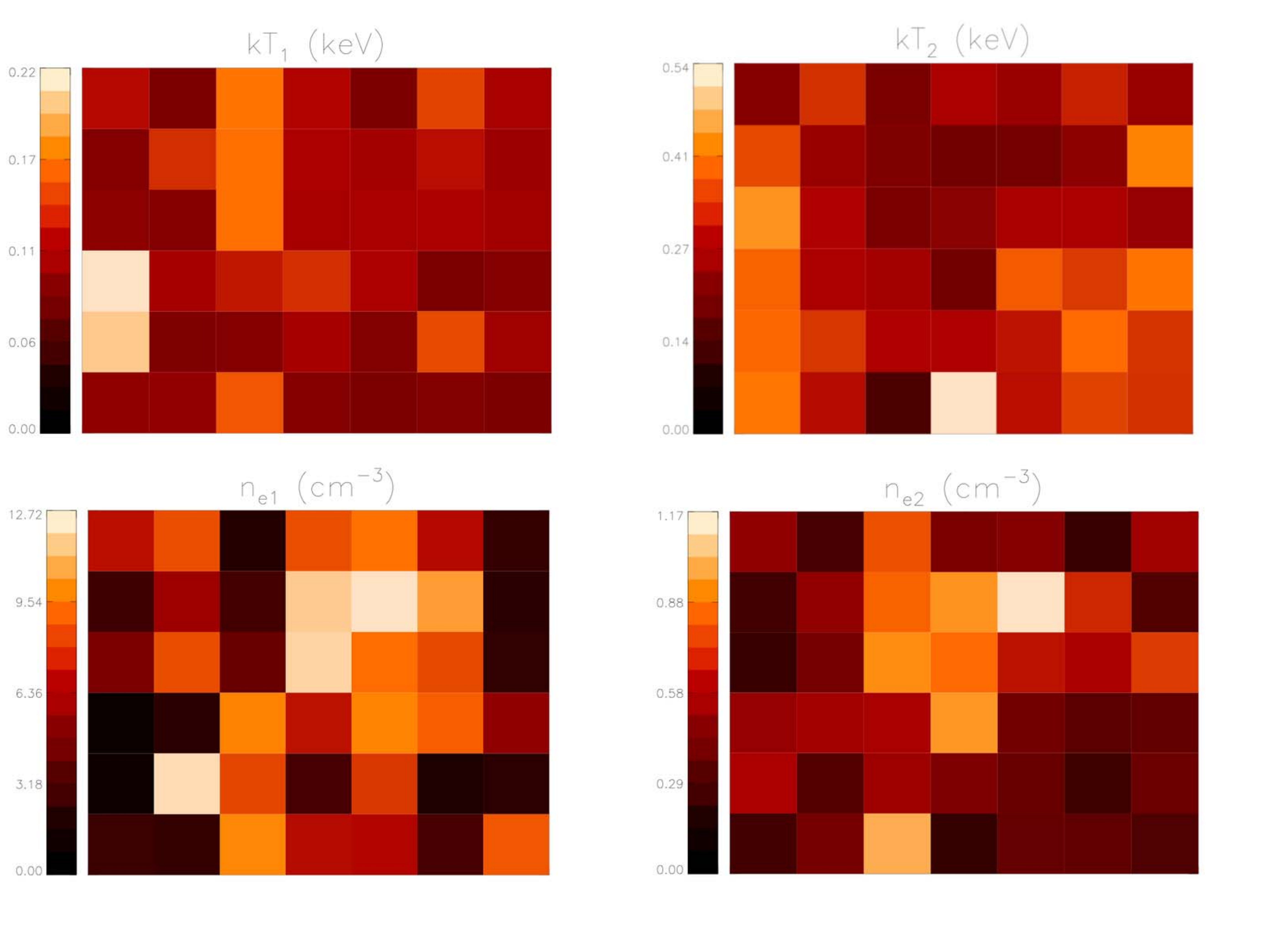}
   \caption{Best fit parameter maps for fits to the bright knot region in the southeast of this observation.  A spectrum was extracted and fit for each of the 42 boxes shown which correspond to the grid of boxes in figure~\ref{grid}.  Fit parameters include temperature and density for each of the two model components.  The lower temperature, typically equilibrium component has parameters $kT_1$ and $n_{e1}$ with $kT_2$ and $n_{e2}$ characterizing the nonequilibrium component.  The vertical color bar gives the corresponding range of values for the colors in the boxes.}
   \label{gridparams}
\end{figure}

Of final note, the bright knot in the southeast stands out even in this highly complex field.  In an attempt to exploit the high spatial resolution of \textit{Chandra}, we extract 42 spectra from this region using a fine grid of boxes, each measuring $27\times22$ pixels (Figure~\ref{grid}).  Given a distance of 540 pc, each box probes a mere $0.034\times0.028$ parsecs of area at the remnant.  The extracted spectra were fit with the same round of models but only the two temperature component models provide an adequate fit.  In general, these regions are well fit with a \textit{vequil} $+$ \textit{vnei} model.  Of the 42 regions, 7 were fit statistically better (via F-test) with a double nonequilibrium model.  Parameter maps of the best fit parameters for $kT$ and $n_e$ over the grid are shown in figure~\ref{gridparams} where $n_e$ is calculated from the fit normalization.  The lower temperature component corresponds to $kT_1$ and $n_{e1}$ and is typically in equilibrium while the higher temperature component has $kT_2$ and $n_{e2}$ and is always in nonequilibrium.  The average error in the equilibrium temperature is $\sim3\%$ and for the nonequilibrium temperature it is $\sim4\%$.  The average errors in density range from 8--11$\%$.  The abundances of C, N, and O were allowed to vary but fixed to each other.  The abundances were typically depleted relative to solar to a similar degree as the other regions (0.3--0.8 times solar) with a few grid boxes around solar values.  The $\chi^{2}_{red}$ has an average and standard deviation of $1.5\pm0.4$ over the grid.

%The temperatures here (0.22--0.37 keV) are consistent with those found in region E.  The lowest temperatures in this region lie along a diagonal that is spatially correlated with the elongated northeastern edge of the bright emission.  This same diagonal also contains the highest values for the ionization parameter.  It is difficult to find substantial spatial correlation in the abundance parameter maps.  However, consistent with other regions, carbon is consistently over abundant when compared to solar and oxygen is consistently depleted.

%We map out the spectral properties of this region by placing a grid of extraction boxes over the entire knot as well as adjacent areas.  A close-up of this region and the extraction grid are shown in figure~\ref{grid}.  There is some consistency between the band ratio map and the false color image in this region but it is not overly obvious.  The bright yellowish knot could be physically congruent to the bright green emission to the northwest.  The band ratio map of the yellow knot shows that it is somewhat overabundant in soft X-ray emission indicating that it could represent a large shockwave/cloud interaction region with quite high density, thus leading to a higher emission measure.  If the green emission is physically tied to the yellow knot, then it could represent a lower density medium that has been reshocked by a reverse shock rebounding from the dense interaction region.  These possibilities are explored in the spectral analysis.

\section{Interpretation}\label{interp}
In general, the extracted spectra are best fit by two temperature component models.  A handful of scenarios may be used to explain this phenomenon.  First, the region may be merely confused due to multiple, physically distinct plasmas along the line of sight.  This effect is expected to become less of an issue toward the rim and more significant for the interior regions.  Second, the region may encompass a high density region and a low density region, both of which have been recently shocked.  The high density region will equilibrate (larger ionization parameter, $n_et$) and cool more rapidly than the low density region, thus leaving two temperature components.  Finally, a similar situation may exist with two distinct density regions, but regions that are causally linked.  Such a situation may occur when the shock wave passes out of the precursor formed cavity and into the cavity wall.  The high density gas in the cavity wall will again equilibrate more rapidly, but the low density region is located interior to the wall and a reverse shock will reshock and additionally heat this interior gas.

In order to distinguish between these scenarios (or develop a new one) we calculate the densities ($n_e$) for each of the model components in all regions.  The results are given in Table~\ref{tbl4}.  The density is calculated using the $norm$ parameter within each component where $norm=[10^{-14}/(4 \pi D^{2})] \int n_{e}n_{H}dV$.  A distance of 540 pc is used for $D$ and it is assumed that $n_e=n_H$.  The volume is calculated using the area of each region multiplied by the depth of the column of remnant gas each region intersects.  This depth is found by assuming the remnant is a filled sphere with radius 1.5$^{\circ}$ or 14 pc and calculating a chord length based on the radial position of each region.  It is assumed that the B regions sample the outermost radii of the sphere.  These assumptions probably lead to the error in volume being the largest contributor to the error in density so that the densities in table~\ref{tbl4} should be treated as approximate.  However, since the volume between components within a region remains constant, the relative density between components in a given region is accurate.

The most evident trend is in the B regions.  In all 5 regions the lower temperature component ($\langle kT \rangle \sim0.10$ keV) exhibits a significant density enhancement - on average, over an order of magnitude higher than the high temperature component ($\langle kT \rangle \sim0.38$ keV).  This result, combined with the location of the B regions at the easternmost edge of the X-ray emission strongly suggests that these B regions trace out the location of the cavity wall.  The high density, low temperature component represents the high density enhancements in the cavity wall itself while the low density, high temperature gas most likely lies just interior to the wall in the precursor formed cavity.  The blast wave has shock heated high density clouds in this wall and reverse shocks from this interaction propagate toward the interior.  The high density gas equilibrates rapidly, cools via emission lines (predominantly \ion{O}{7}) and has a lower temperature than the lower density interior which has been reshocked and takes longer to cool.

%The equilibrium nature of the B region gas can be used to lend a few clues to the time evolution of this section of the Cygnus Loop.  In each of the B regions we may be sampling the rapidly equilibrated cavity wall and an interior plasma that has been reshocked by a reverse shock.  

It may be expected that the interior lower density gas is in nonequilibrium, but that is only true if it has been very recently shocked, i.e. $t$ and $n_e$ are sufficiently low to give a low ionization parameter.  However, in all B regions the high temperature component can never be fit with a nonequilibrium plasma and any attempts drive the ionization parameter to its upper limit, which signifies equilibrium and therefore a longer timescale since being shocked.  Solving the Rankine-Hugoniot relations for a strong shock and $\gamma=5/3$ gives the post-shock temperature as a function of shock velocity, $T=(3/16)\mu$$v^2/k$ with $\mu=1.2$.  This can be easily applied to the cavity wall where the average shock velocity in this high density gas is 209 km/sec.  If we then assume a Sedov solution the time since shock heating can be calculated as $t=(2/5)R/v$ where $R$ is the radial shock location, 14 pc.  While this is an inaccurate assumption in the cavity wall it will provide an upper limit to the time since the shock velocity in the dense gas is lower than the Sedov velocity.  This upper limit is $\sim$26,000 years.  Ideally we could use the shock velocity combined with visible spatial offsets between the shocks in the cloud and the reverse shock to estimate a propagation timescale, but the resolution and/or count rate is not present in this observation.  The situation for the interior gas is even more complicated if we assume it has been shocked twice, once by the initial supernova blast wave and again by a reverse shock.  The contribution from each of these shocks to the temperature of the gas is uncertain.  Therefore, a similar calculation will overestimate the shock velocity.  Even so, the high temperature component in these regions is still best fit with an equilibrium plasma which suggests high densities and/or long timescales since shock heating.  Therefore, this crude age estimation, the more precise velocity calculation, and the large ionization timescales suggest that the B regions have equilibrated due to their high density and possibly also due to a significant time since being shocked.  
 
%while the average velocity in the low density gas is nearly twice as high, 395 km/sec.  The time since shock heating can then be calculated as $t=(2/5)R/v$ where $R$ is the radial shock location, 14 pc.  This gives lifetimes of ~26,000 years for the low temperature gas and ~14,000 years for the high temperature gas.  These figures may trend toward the high end given the Sedov assumption, but nevertheless they are suggestive of the interior gas being shocked at a later time than the high density cavity wall, probably by a reverse shock from the wall.  Using the difference in shock times and the interior shock velocity shows that the reverse shock should have been carried $\sim$5 pc, easily capable of encompassing the B regions.  Therefore, the B regions have equilibrated not only due to their high density, but also due to a considerable time since being shocked.  These shock times and densities can be used to approximate the expected ionization parameters for the B region components.  These values are large as expected and are calculated to be $n_et\sim4\times10^{12}$ cm$^{-3}$ s and $n_et\sim2\times10^{11}$ cm$^{-3}$ s for the low temperature and high temperature components, respectively.

\begin{figure} [htbp]
   \centering
   \epsscale{1.0}
%   \plottwo{f4a.eps}{f4b.eps}
%	 \plotone{f4.eps}
   \includegraphics[width=6.0in,height=3.115in]{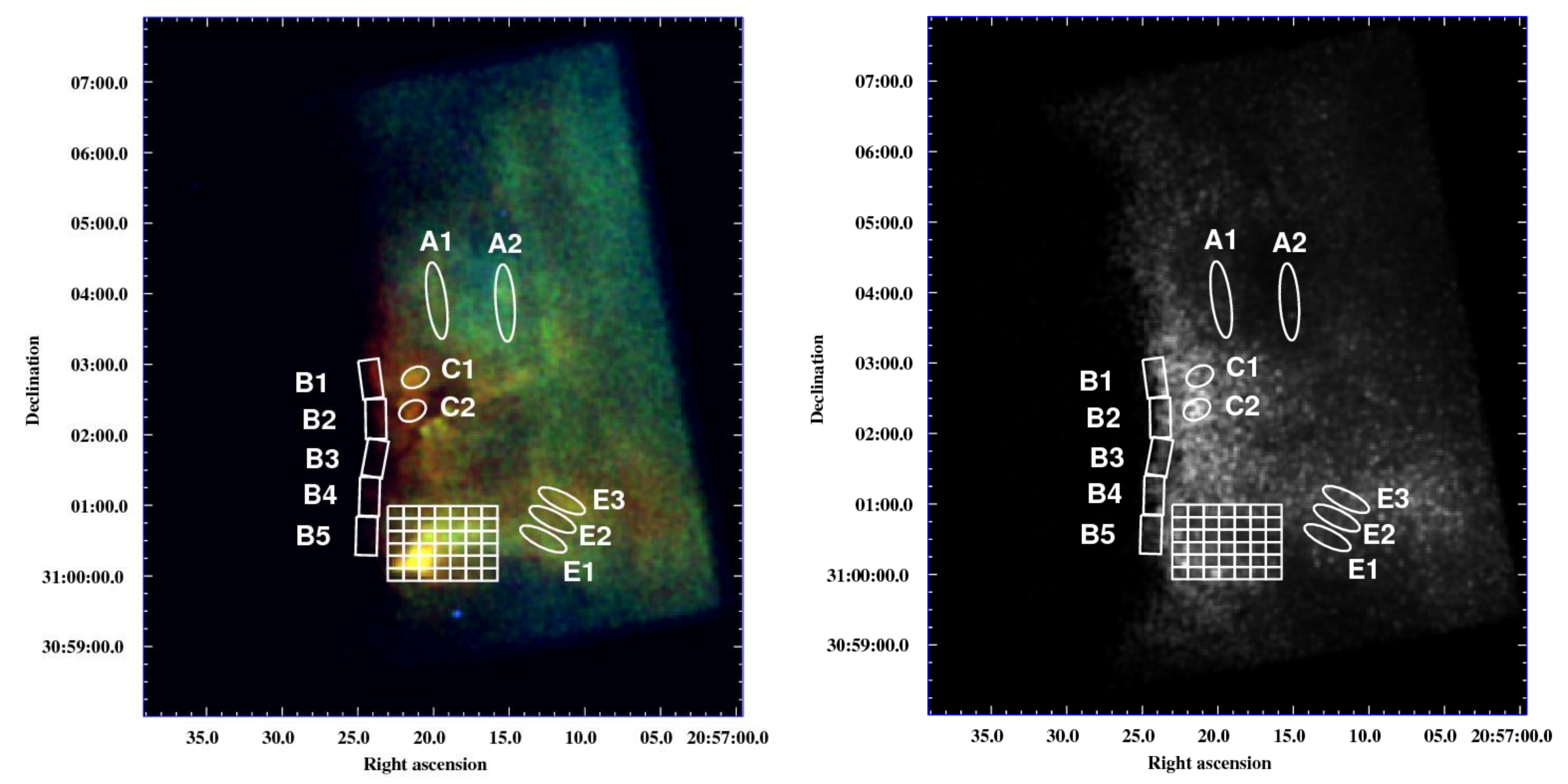}
   \caption{Spectral extraction regions are overlaid on the band ratio map and the false color image.  The regions are labeled are on the right and were chosen based on the morphology evident in these images.}
   \label{regions}
\end{figure}

In other regions the situation is complicated by their interior locations and presumed line of sight confusion.  For instance, the A regions exhibit similar properties to the B regions with high densities typifying the low temperature component.  This effect is more pronounced in region A1 given the more distinct temperature difference between components in comparison to region A2.  However, given the location of these regions it is more likely that line of sight confusion leads to multiple components rather than probing a single distinct cavity wall/shock interaction as with the B regions.  Inspection of figure~\ref{regions} shows that A1 is dominated by hard emission, but may lie close enough to the rim of the remnant to pick up some low temperature cavity wall emission along the line of sight.  This is supported by the presence of a very high density, low temperature component in the A1 fit.  Also, the presence of soft x-rays is somewhat evident in both the false color and ratio maps.  In region A2 though there is a distinct absence of soft X-ray emission in figure~\ref{regions}.  The A2 line of sight is located interior to A1 and therefore probes through a much larger column of the high temperature gas, which dominates as shown by the best fit temperatures of $kT\sim$0.23--0.30 keV.  If we assume that the low temperature component is completely contained within a shell of thickness comparable to a B region, and use a radius of 14 pc for the location of the cavity wall, then we can calculate the relative contributions of the high and low temperature components in each of these A regions.  Region A1 probes a total shell volume of 0.015 pc$^3$ and interior volume of 0.12 pc$^3$ while the A1 line of sight contains a shell volume of 0.011 pc$^3$ and interior volume of 0.16 pc$^3$.  This estimates that region A2 contains approximately twice as much interior, high temperature emission relative to low temperature emission.  The lower density, high temperature plasmas of A2 are consistent with nonequilibrium conditions for this interior region.  Furthermore, the interior plasmas of these A regions display significant emission from higher energy lines, most likely \ion{Fe}{17} and \ion{Ne}{9}, also consistent with the high temperatures in these regions.  Therefore, while the A regions probably encompass some cavity wall emission, some reverse shock emission, some singly shocked low density hot emission, or some combination of all of the above, it is unclear if the temperature/density components are related given that these regions probe an interior sightline that is prone to confusion.

The C regions follow a similar trend as the A regions.  C2 shares properties of A1; the line of sight probes both a low temperature, equilibrium plasma near the edge of emission and hotter ineterior gas as well.  The presence of the low temperature component is quite evident upon examination of the ratio map.  C2 outlines the brightest region in this map so it is not surprising that there is a significant soft component.  This component is probably due to a density enhancement in the cavity wall ($n_e=5.6$), but possibly located in the foreground, toward the observer in comparison to the B regions, assuming a spherical shape to the remnant.  However, the exact location of the soft emission along the line of sight is complicated by the high degree of complexity in this area as seen in the raw data and color image.  Region C1 gives no further clues either given the distinctly different fit parameters in comparison to C2.  This particular line of sight probes mostly higher temperature gas.  It is quite bright in the raw data, but shows no significant abudance of soft X-ray emission in the ratio map, especially in comparison to C2.  The lack of soft X-rays leads to the relatively higher fit temperatures ($kT=0.19$ keV versus 0.10 keV on average for soft X-ray components) while the bright emission is consistent with the high density, i.e. large emission measure, of this region.  Region C1 is the only example of a high density, high temperature, nonequilibrium plasma in any of the extracted regions.  If this density enhancement is in nonequilibrium then it must have been recently shocked.  Using the ionization parameter and the calculated density show that this plasma was most likely shocked within the last 1000 years.  It is possible that this cloud traces out where the blast wave is interacting with the cavity wall but at a larger radius than the B regions, thus leading to the more recent shock time.  If this is the case, then the cloud must be located considerably in the foreground relative to B.  Therefore, the C regions are again confused by line of sight projection yet show the presence of probable cavity wall emission in addition to the hotter, lower density, interior emission.

The picture is not so clear in the E regions.  These extractions are even farther toward the interior than the A regions, but seem to have simpler fits.  For instance, a single temperature equilibrium plasma is the only reasonable fit obtained for region E3.  One might expect that a multiple component fit at high temperature and nonequilibrium may be favored.  However, these regions were chosen because they sample the soft X-ray extension that protrudes westward from the edge of the remnant (see Figure~\ref{regions}).  It may be the case that region E3 looks directly at a dense cavity wall cloud, which obscures the view into the hotter, low density interior.  The calculated density of 3.0 cm$^{-3}$ is not overly dense, but still a few--several times denser than interior densities seen in other regions.  Region E2 on the other hand is well fit with a two component model containing both a low temperature equilibrium plasma with a high temperature nonequilibrium plasma.  This again may be indicative of cavity wall emission in a dense medium that has equilibrated and cooled combined with interior plasma that is at a higher temperature (possibly reshocked) and lower density leading to nonequilibrium conditions.  However, the calculated densities are not as clear cut as they are in previous regions.  The low temperature component for E2 is at approximately the same density as the high temperature component thus lending no resolution to the line of sight confusion.  Inspection of the color map shows that while E3 is situated in the midst of soft emission, E2 is sampling a region with a mix of low and higher energy emission with no obvious distinction.  Finally, E1 is even further afield from the soft emission of E3 than E2 and appears to be dominated by harder X-rays.  However, the best fit spectrum is only marginally hotter than E3 at 0.17 keV and at nearly the same density.  The main difference between these regions is that E1 appears to be in nonequilibrium and therefore more recently shocked relative to E3.  Perhaps E1 is a medium density cloud that has been reshocked by a reverse shock originating in the denser cavity wall.  While the E regions provide interesting hints into the morphology and possible interactions in this section of the Cygnus Loop, it is impossible to state anything definitively given the confusion inherent to these interior regions.

The grid analysis of the bright southeast knot shown in Figure~\ref{gridparams} has presented some intriguing results as well.  Inspection of the equilibrium maps of $kT_1$ and $n_{e1}$ reveal that they appear almost as negatives of one another.  As expected, the highest densities correspond to areas of bright emission with density falling off toward the edges of the grid.  These dense pockets have slowed the shock considerably more than the surrounding shock regions.  The slower shock velocity results in correspondingly lower post-shock temperature.  The densities calculated here are similar to those found in the cavity wall along the B regions.  This bright area of emission could be an indicator of a protrusion into the cavity wall or mark a particularly large, dense cloud along the cavity wall slightly in the foreground relative to the B regions, or a combination of the two.  Outside of the bright emission the color map contains harder emission which is consistent with the higher temperature fits in these grid squares.  These regions could be located outside of the dense cloud and probe interior gas in the background, along the line of sight.  The brightest knot in the southeast, which displays high density, is located close to the edge of the X-ray emission.  In contrast to region B which appears to follow an $\sim$0.43 pc vertically elongated section of the cavity wall, this particular emission seems to be located in a more discrete cloud.  This cloud could possibly be oriented along the line of sight, which may help in explaining the intensity of emission and the apparent indentation in the X-ray emission at this point.  Previous studies of the XA region have explained the X-ray emission as arising from a single, large protrusion which is responsible for the entire field of view studied here.  However, the fine spatial resolution of \textit{Chandra} has uncovered much smaller blast wave/cloud interactions, possibly as small as a single grid square, 0.035 pc $\times$ 0.028 pc.

\section{Conclusions}\label{conc}
The high spatial resolution of \textit{Chandra} has given us insight into the structure of the complicated XA region.  Specifically, the chain of B regions appear to follow the location of the blast wave interaction with the cavity wall.  These B regions are located at the very edge of X-ray emission and are dominated by soft X-ray emission.  Spectral analysis has shown low temperature gas at densities an order of magnitude higher than more recently shocked high temperature gas.  Shock heated gas in the high density cavity wall has equilibrated rapidly and is efficiently cooling through \ion{O}{7} emission.  A reverse shock has propagated from the initial supernova blast wave interaction with the cavity wall and has reheated the interior gas.  Furthermore, the time since interacton of the reverse shock with the interior gas has been long enough for the plasma to equilibrate as well.  Given this interpretation, the precursor wind blew a cavity that is $\sim$ 14 pc in radius.

The identification of the cavity wall is simpler and more robust than identifying structure interior to the edge.  While the color map shows a general trend of soft X-ray emission on the exterior with harder emission toward the interior, this trend is not obvious in discrete extractions.  Regions interior to the B regions have line of sight confusion and significant chord length through the spherically shaped remnant.  Some of these regions show strong signs of including the cavity wall (high density, low temperature components), but the interpretation is less obvious given that these regions are not located at the rim of emission.  Furthermore there is a mix of temperatures, densities, and ionization states with no strong trends.  However, we feel that this is a significant finding given that previous studies support a much simpler morphology whether it be a single finger-like protrusion into this region or a few clumps.  We would argue that the fine spatial resolution here has uncovered a situation where any region removed from the edge of emission will have significant line of sight confusion and may or may not include cavity wall emission, interior emission, interior reverse shock emission, or multiple components of each.  It is not to say that any given region is physically disconnected from another since it is definitely the case that the X-ray morphology is arising from a global interaction of the blast wave with a precursor formed cavity wall.  However, the intricate details of the X-ray emission and therefore the physical parameters of the plasma can vary over the smallest scales.  This effect is evident in the gridded extraction region around the bright knot of emission in the southeast.  The calculated densities in the grid show the presence of an isolated blast wave/cloud interaction possibly dominated by only a few grid sections.  These data display a significant increase in resolution over the previously highest resolution study \citep{Zhou} where extraction regions are larger than the entire grid.

The Cygnus Loop is clearly complex both spatially and spectrally.  \textit{Chandra} can image fine structures indicative of distinct interaction regions, but the nearly broadband spectra are lagging in quality.  The next step in understanding the intricacies inherent to the interaction of this supernova with the ISM is to obtain high quality spectral data.  Higher spectral resolution is required to constrain the parameters of the models and test their validity.  Grating instruments onboard existing X-ray telescopes are capable of high spectral resolution but are designed for point sources.  The diffuse nature of galactic supernova remnants cause current grating observations to be signal limited with blurred spectral resolution.  To increase understanding of these objects future instruments will need to be capable of performing high efficiency and high spectral resolution observations of diffuse soft X-ray sources.

\section{Acknowledgments}\label{ack}
The authors would like to acknowledge internal funding initiatives at the University of Iowa for support of this work.  The data used here were obtained from the Chandra Data Archive.

\clearpage

%table 1

\begin{table*}[tbp]
\caption{Best Fit Parameters for Regions A and B.}\label{tbl1}
\scalebox{0.65}{%
\centering
\begin{minipage}{\textwidth}

\begin{tabular}{l c c c c c c c	c	c	c	c}
%\multicolumn{8}{c}{\textsc{Table 1}}\\
%\multicolumn{8}{c}{Equilibrium Plasma Spectral Fit Parameters}\\
	\hline
	\hline
	Region\footnote[0]{Note.---Abundances are shown relative to solar with solar values equal to 1.} & {$kT_1$} & {CNO$_1$}\footnote[1]{All B regions vary only O.  C and N are set to solar values. CNO$_1$ and CNO$_2$ are independent in A regions but fixed equal in B regions.} & {$\tau_1$} & {norm$_1$} & {$kT_2$}	&	{CNO$_2$}	&	{$\tau_2$}	&	{norm$_2$}	&	N$_{cts}$\footnote[2]{Total counts per extraction region}	&	{$\chi^2$/dof}	&\\
	&	(keV) & & ($10^{11} cm^{-3} s$) & ($10^{-4}A$\footnote[3]{Normalization parameter, where $A=[10^{-14}/(4 \pi D^{2})] \int n_{e}n_{H}dV$, D is the distance to the Cygnus Loop and the integral is the emission measure.}) & (keV) & &	($10^{11} cm^{-3} s$)	&	($10^{-4}A$)	&\\
	\hline
	A1 & 0.0861$^{+0.0004}_{-0.0008}$ & 0.109$\pm$0.006 & n/a & 280$\pm$10 & 0.236$\pm$0.003 & 0.69$\pm$0.04 & n/a	&	2.54$\pm0.08$	&	7100	&	49.4/35\\
	A2 & 0.232$^{+0.005}_{-0.006}$ & 0.6$\pm$0.1 & 0.0033$^{+0.0002}_{-0.0003}$ & 29$\pm$4 & 0.299$\pm$0.003 & 0.61$\pm$0.02 & 0.60$\pm$0.05	&	1.94$\pm0.04$	&	9391	&	51.6/40\\
	B1 & 0.095$\pm$0.001 & 0.59$\pm$0.03 & n/a & 12.8$\pm$0.7 & 0.37$^{+0.05}_{-0.04}$ & 0.59$\pm$0.03 & n/a	&	0.07$\pm0.01$	&	1275	&	24.1/13\\
	B2 & 0.085$\pm$0.001 & 0.51$\pm$0.03 & n/a & 25$\pm$1 & 0.26$\pm$0.02 & 0.51$\pm$0.03 &	n/a	&	0.17$\pm$0.03	&	1176	&	21.6/13\\
	B3 & 0.096$\pm$0.001 & 0.82$\pm$0.06 & n/a & 5.9$\pm$0.4 & 0.27$\pm$0.02 & 0.82$\pm$0.06 & n/a	&	0.12$\pm$0.02	&	961	&	21.7/13\\
	B4 & 0.116$\pm$0.002 & 0.55$\pm$0.04 & n/a & 4.3$\pm0.3$ & 0.38$^{+0.09}_{-0.08}$ & 0.55$\pm$0.04 & n/a	&	0.03$\pm$0.01	&	951	&	17.1/11\\
	B5 & 0.123$\pm$0.002 & 0.79$\pm$0.05 & n/a & 2.7$\pm$0.2 & 0.6$\pm$0.1 & 0.79$\pm$0.05 & n/a	&	0.037$\pm$0.009	&	1061	&	26.2/14\\
	\hline
\end{tabular}
\end{minipage}}
\end{table*}

\clearpage

%table 2

\begin{table*}[tbp]
\caption{Best Fit Parameters for Region C.}\label{tbl2}
\scalebox{0.65}{%
\centering
\begin{minipage}{\textwidth}

\begin{tabular}{l c c c c c c c	c	c	c	c	c}
%\multicolumn{8}{c}{\textsc{Table 1}}\\
%\multicolumn{8}{c}{Equilibrium Plasma Spectral Fit Parameters}\\
	\hline
	\hline
	Region\footnote[0]{Note.---Abundances are shown relative to solar with solar values equal to 1.} & {$kT_1$} & {$\tau_1$} & {norm$_1$} & {$kT_2$}	&	{$\tau_2$}	&	{norm$_2$}	&	C	&	N	&	O	&	N$_{cts}$\footnote[1]{Total counts per extraction region}	&	{$\chi^2$/dof}	&\\
	&	(keV) & ($10^{11} cm^{-3} s$) & ($10^{-4}A$\footnote[2]{Normalization parameter, where $A=[10^{-14}/(4 \pi D^{2})] \int n_{e}n_{H}dV$, D is the distance to the Cygnus Loop and the integral is the emission measure.}) & (keV) &	($10^{11} cm^{-3} s$)	&	($10^{-4}A$)	&\\
	\hline
	C1 & 0.179$\pm$0.006 & 0.06$^{+0.03}_{-0.01}$ & 0.66$\pm$0.07 & 0.191$\pm$0.002 & 0.0043$\pm$0.0007 & 38$\pm$2 & $<$7.5	&	0.7$\pm$0.2	&	0.81$\pm$0.05	&	2576	&	27.4/16\\
	C2 & 0.107$\pm$0.001 & n/a & 11.7$\pm$0.5 & 0.37$^{+0.09}_{-0.08}$ & n/a & 0.04$\pm$0.01 & 1.0	&	1.0	&	0.51$\pm$0.03	&	1597	&	25.8/18\\
	\hline
\end{tabular}
\end{minipage}}
\end{table*}

\clearpage

%table 3

\begin{table*}[tbp]
\caption{Best Fit Parameters for Region E.}\label{tbl3}
\scalebox{0.65}{%
\centering
\begin{minipage}{\textwidth}

\begin{tabular}{l c c c c c c c	c	c	c	c}
%\multicolumn{8}{c}{\textsc{Table 1}}\\
%\multicolumn{8}{c}{Equilibrium Plasma Spectral Fit Parameters}\\
	\hline
	\hline
	Region\footnote[0]{Note.---Abundances are shown relative to solar with solar values equal to 1.} & {$kT_1$} & O	&	Ne	&	Fe	&	{$\tau_1$} & {norm$_1$} & {$kT_2$}	&	{norm$_2$}	&	N$_{cts}$\footnote[1]{Total counts per extraction region}	&	{$\chi^2$/dof}	&\\
	&	(keV) & &	&	&	($10^{11} cm^{-3} s$) & ($10^{-4}A$\footnote[2]{Normalization parameter, where $A=[10^{-14}/(4 \pi D^{2})] \int n_{e}n_{H}dV$, D is the distance to the Cygnus Loop and the integral is the emission measure.}) & (keV) &	($10^{-4}A$)	&\\
	\hline
	E1 & 0.174$\pm$0.001 & 0.217$\pm$0.009 & 0.35$\pm$0.05 & 0.8$\pm$0.1 & 2.5$^{+0.5}_{-0.4}$ & 10.5$\pm$0.3 & n/a	&	n/a	&	4333	&	50.7/28\\
	E2 & 0.198$\pm$0.002 & 0.12$\pm$0.01 & 0.32$\pm$0.07 & 1 & 0.8$\pm$0.1 & 4.6$\pm$0.2 & 0.146$\pm$0.003	&	2.4$\pm$0.1	&	3910	&	55.3/27\\
	E3 & 0.166$\pm$0.001 & 0.25$\pm$0.01 & 0.41$\pm$0.07 & 1.6$\pm$0.2 & n/a & 12.1$\pm$0.3 & n/a	&	n/a	&	4330	&	45.8/31\\
	\hline
\end{tabular}
\end{minipage}}
\end{table*}

\clearpage

%table 4

\begin{table*}[tbp]
\caption{Electron Density and Shock Velocity.}\label{tbl4}
\scalebox{0.8}{%
\centering
\begin{minipage}{\textwidth}

\begin{tabular}{l c c c	c	c}
%\multicolumn{8}{c}{\textsc{Table 1}}\\
%\multicolumn{8}{c}{Equilibrium Plasma Spectral Fit Parameters}\\
	\hline
	\hline
	Region\footnote[0]{Note.---The first density listed corresponds to the lower T component for regions with two components.} & $n_e$ (cm$^{-3}$) &	Vol. (pc$^3$)	& $v$ (km/sec)\\
	\hline
	A1 & 13, 1.2	&	0.041	&	192, 317\\
	A2 & 3.4, 0.8	&	0.058	&	315, 357\\
	B1 & 9.5, 0.7	&	0.012	&	201, 397\\
	B2 & 13, 1.0	&	0.012	&	190, 332\\
	B3 & 6.4, 1.0	&	0.012	&	202, 339\\
	B4 & 5.4, 0.4	&	0.012	&	222, 403\\
	B5 & 4.4, 0.6	&	0.012	&	229, 506\\
	C1 & 1.4, 10	&	0.010	&	276, 285\\
	C2 & 5.6, 0.4	&	0.010	&	272\\
	E1 & 2.8	&	0.035	&	250, 291\\
	E2 & 1.4, 1.8	&	0.035	&	192, 317\\
	E3 & 3.0	&	0.035	&	266\\
	\hline
\end{tabular}
\end{minipage}}
\end{table*}

\clearpage

\clearpage

%figure 2:

\clearpage

%figure 3:

\clearpage

%figure 4:

\clearpage

%figure 5:

\clearpage

%figure 6:

\clearpage

%figure 7:

\clearpage

%figure 8:
%\begin{figure} [htbp]
%   \centering
%   \epsscale{0.5}
%   \plotone{f8.eps}
%%   \includegraphics[width=3.0in,height=4.84in]{figs/Bshocks.pdf}
%   \caption{Contrast enhanced close-up of the shock front.  Intricate structure is evident at the smalles scales including a possible low density "`breakout"' region.  Shocks indicated by arrows correspond to B1-B4 with B5 on the breakout and B6 and B7 at the bottom.}
%   \label{Bshocks}
%\end{figure}

%figure 9:
%\begin{figure} [htbp]
%   \centering
%   \epsscale{1.0}
%   \plotone{f9.eps}
%%   \includegraphics[width=6.0in,height=3.923in]{figs/xoptical2.png}
%   \caption{X-ray flux contours (white lines) overlayed on optical images.  The green image on the left is [\ion{O}{3}] with H$\alpha$ in red on the right.}
%   \label{xoptical}
%\end{figure}

\end{document}